\documentclass{vgtc}                          

\graphicspath{{figures/}{pictures/}{images/}{./}} 

\usepackage{times}                     

\usepackage{tabu}                      
\usepackage{booktabs}                  
\usepackage{lipsum}                    
\usepackage{mwe}                       

\usepackage{mathptmx}     
\usepackage{eucal}
\usepackage{amssymb}
\usepackage{comment}
\usepackage{amsthm}
\usepackage{amsmath}
\usepackage[linesnumbered,ruled]{algorithm2e}
\usepackage{float}
\usepackage{caption}
\usepackage{subcaption}

\newcommand{\etal}{\textit{et al.}}

\newcommand{\new}[1]{\textcolor{black}{#1}}

\onlineid{0}

\vgtccategory{Research}

\vgtcinsertpkg

\title{Multi-scale Cycle Tracking in Dynamic Planar Graphs}

\author{Farhan Rasheed\thanks{e-mail:\{farhan.rasheed$|$emma.nilsson$|\newline$talha.bin.masood$|$ingrid.hotz\}@liu.se\\ Affiliation:\ Link\"oping University}
\and Abrar Naseer\thanks{e-mail: abrarnaseer@iisc.ac.in\\ Affiliation:\ Indian Institute of Science}
\and Emma Nilsson\footnotemark[1]
\and Talha Bin Masood\footnotemark[1]
\and Ingrid Hotz\footnotemark[1]
}

\authorfooter{

\item
 Farhan Rasheed. E-mail: farhan.rasheed@liu.se.
\item
 Emma Nilsson. E-mail: emma.nilsson@liu.se.
\item
 Talha bin Masood. E-mail: talha.bin.masood@liu.se.
\item
 Ingrid Hotz. E-mail: ingrid.hotz@liu.se.
}

\teaser{
  \centering
  \includegraphics[width=\textwidth]{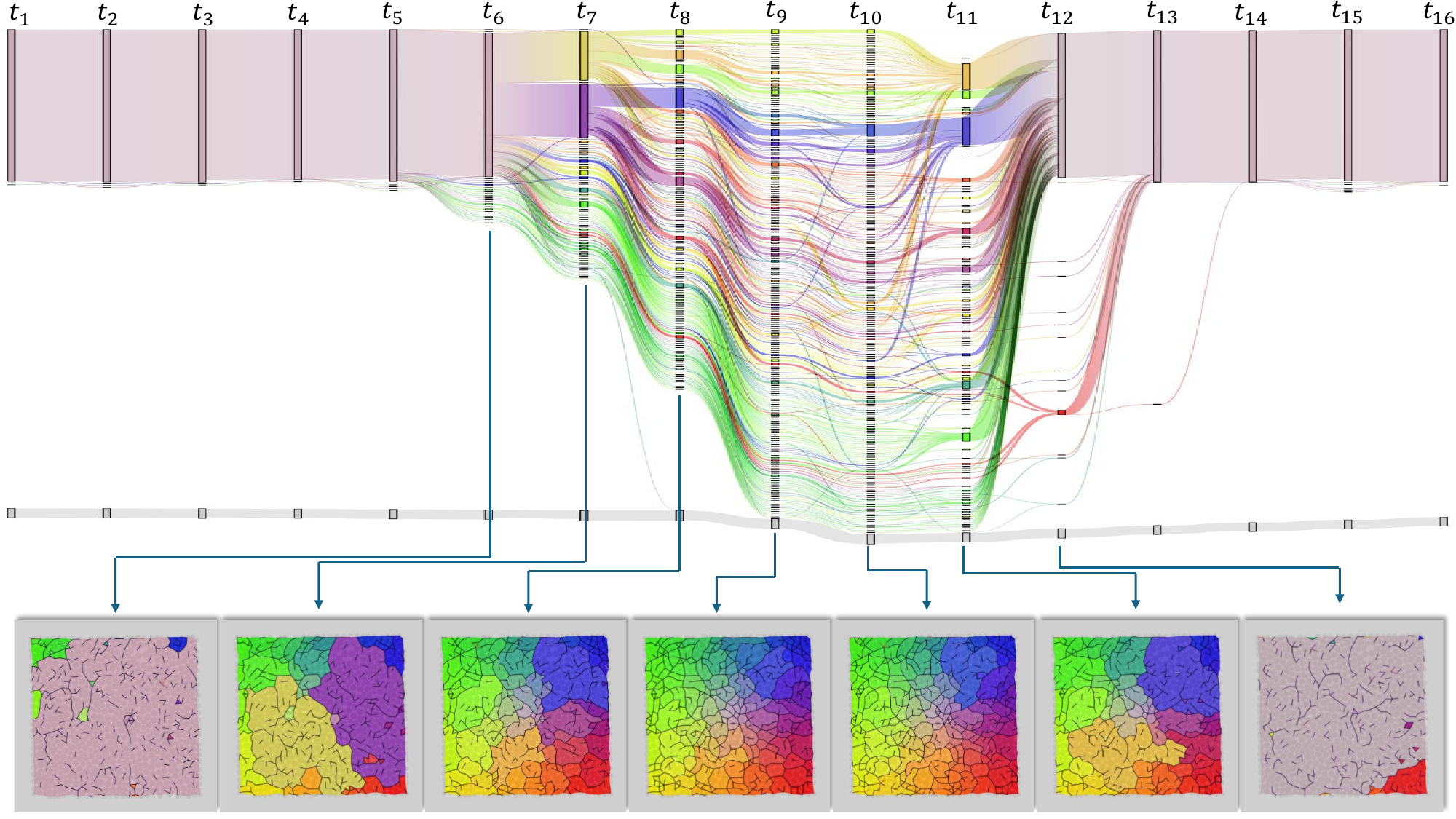}
  \caption{{Experiment B: \new{The tracking graph illustrates the transition of a granular material system to a rigid state ($t_6$ to $t_{12}$) under the influence of external stress. It is observed by looking at the cycle evolution in the underlying force network within the granular system. As the external stresses increase by increasing the load on the boundaries of the system, the cycles start splitting into smaller cycles while with the decrease in the external stress, the cycles start to merge. This splitting and merging of the cycles can be nicely observed in the tracking graph and in the rendering of the granular system subjected to external stresses, shown at the bottom. Each cycle within the force network corresponds to a node in the tracking graph.}}}
  \label{fig:teaser}
}

\abstract{
This paper presents a nested tracking framework for analyzing cycles in 2D force networks within granular materials. These materials are composed of interacting particles, whose interactions are described by a force network. Understanding the cycles within these networks at various scales and their evolution under external loads is crucial, as they significantly contribute to the mechanical and kinematic properties of the system. Our approach involves computing a cycle hierarchy by partitioning the 2D domain into segments bounded by cycles in the force network. We can adapt concepts from nested tracking graphs originally developed for merge trees by leveraging the duality between this partitioning and the cycles. We demonstrate the effectiveness of our method on two force networks derived from experiments with photoelastic disks.

} 

\keywords{Tracking cycles, force network, granular materials, persistence homology, force loops, nested tracking.}

\begin{document}

\maketitle

\newpage
\section{Introduction}
\label{sec:intro}
In this paper, we introduce a method for temporal tracking, analysis, and visualization of multi-scale cycles in planar graphs representing force networks in two-dimensional granular materials.
Granular materials, such as sand, grains, and powders, consist of large collections of discrete particles and appear in a wide range of applications. Depending on external loading conditions, these materials can exhibit characteristics akin to solids, liquids, or gases. It is assumed that microscopic inter-particle interactions largely contribute to the large-scale properties of granular materials. However, the relationship between the properties observed at different scales is not yet fully understood. At the microscopic scale, the inter-particle contacts and forces are observed and can, respectively, be interpreted as networks~\cite{papadopoulos2018network}, whereas, at the macroscopic scale, the material's overall response to external loading is of interest. 
For two-dimensional granular materials, Camout et al.~\cite{Cambou2016} introduced an intermediate scale forming representative volume elements~(RVEs), known as the \emph{mesoscale}. These RVEs are then utilized to aggregate material properties, to provide a more comprehensive understanding of the material's behavior~\cite{Yang2021}. In their work, the RVEs are defined as ``loops'' in the contact network contributing to the mechanic and kinematic properties of the system.

To better understand systems composed of granular material, two-dimensional systems are studied using \emph{experiments with photoelastic disks}, which are the focus of this paper. Thereby the evolution of force and contact networks in response to different loading conditions is observed. The experimental setup includes a collection of photoelastic disks arranged on a near-frictionless table enclosed by a flexible rack. Initially, the disks are in an unjammed state, meaning they do not exhibit any yield stress. During the experiment, the system is gradually and quasi-statically compressed by moving the rack until it is subjected to high stress. The system is then gradually released. This compression and release process is repeated several times, each time starting with the same initial conditions to generate an ensemble of experimental runs.
The photoelastic disks change optical properties under mechanical stress, which can be observed through polarizing filters~\cite{Zadeh2019}. This allows the reconstruction of the force network under different loading conditions. As a result, these experiments produce ensembles of time series of force networks. 
The main goal of these experiments is to observe and quantify structural changes in the force network. This entails keeping track of isolated non-participating particles as well as cycles formed by force chains, which helps to identify the stability of internal structures. In the next step, they also want to be able to compare different experiments to each other and identify common patterns.

In this paper, we examine the temporal aspect of these experiments by tracking cycles throughout the experiments on different scales. These cycles extend the concept of contact loops introduced by Cambou et al.~\cite{Cambou2016} to define RVEs. They provide a multi-scale decomposition of the domain into areas bounded by ``force loops''~\cite{rasheed2023multi}, enclosing particles that do not participate in the force propagation at a given force level. Utilizing a topological framework for analyzing weighted planar graphs to extract the loops, we will refer to the force networks as graphs and the loops as cycles. By employing dual filtration of the force network, \new{introduced in~\cite{rasheed2023multi}}, we obtain a cycle hierarchy that can be represented in the form of a merge tree. This tree forms the basis for multi-scale cycle tracking.
To do so, we build on the concept of nested tracking graphs~\cite{lukasczyk2017nested} that track superlevel set components for different levels simultaneously. Different from the original work, we track a decomposition of the domain defined by the cycles, where each segment is given as a set of triangles of a time-dependent triangulation. Moreover, we allow different levels for the different time steps, adapted to the current range of force values.

For the analysis of the tracks, we provide an interactive visualization environment that shows the tracking graph linked to a representation of the domain partitioning and tree structures, if required. The tracking graph visualizes a one-dimensional projection of the partitioning for each time step and their connectivity using a Sankey diagram~\cite{Kennedy1898} that encodes the changes of the cycles from one time step to the next. Depending on the user's interest, the projection preserves certain properties, such as the range of the cycles. All these components are linked together in Inviwo~\cite{Jonsson2020b} and provide a powerful framework for exploring the properties of tracking graphs.
We demonstrate our framework for two different experiments with photoelastic disks.

\new{
We summarize the contribution of this paper as follows:\begin{itemize}
 \item We present an efficient framework for tracking cycles in planar graphs at multiple scales using dual filtration~\cite{rasheed2023multi}. 
 \item We adapt the nested tracking method for dynamic planar embedded graphs providing domain partitioning. 
\item We propose color strategies for tracking graphs to enhance context and encode additional information: (1) a 2D colormap to emphasize spatial information of cycles, and (2) node coloring based on parent information.
\end{itemize}}
\section{\textbf{Granular Materials and Photoelastic experiment}}
\label{sec:data}
\begin{figure}[h!]
    \centering
    {    \includegraphics[width=\linewidth]{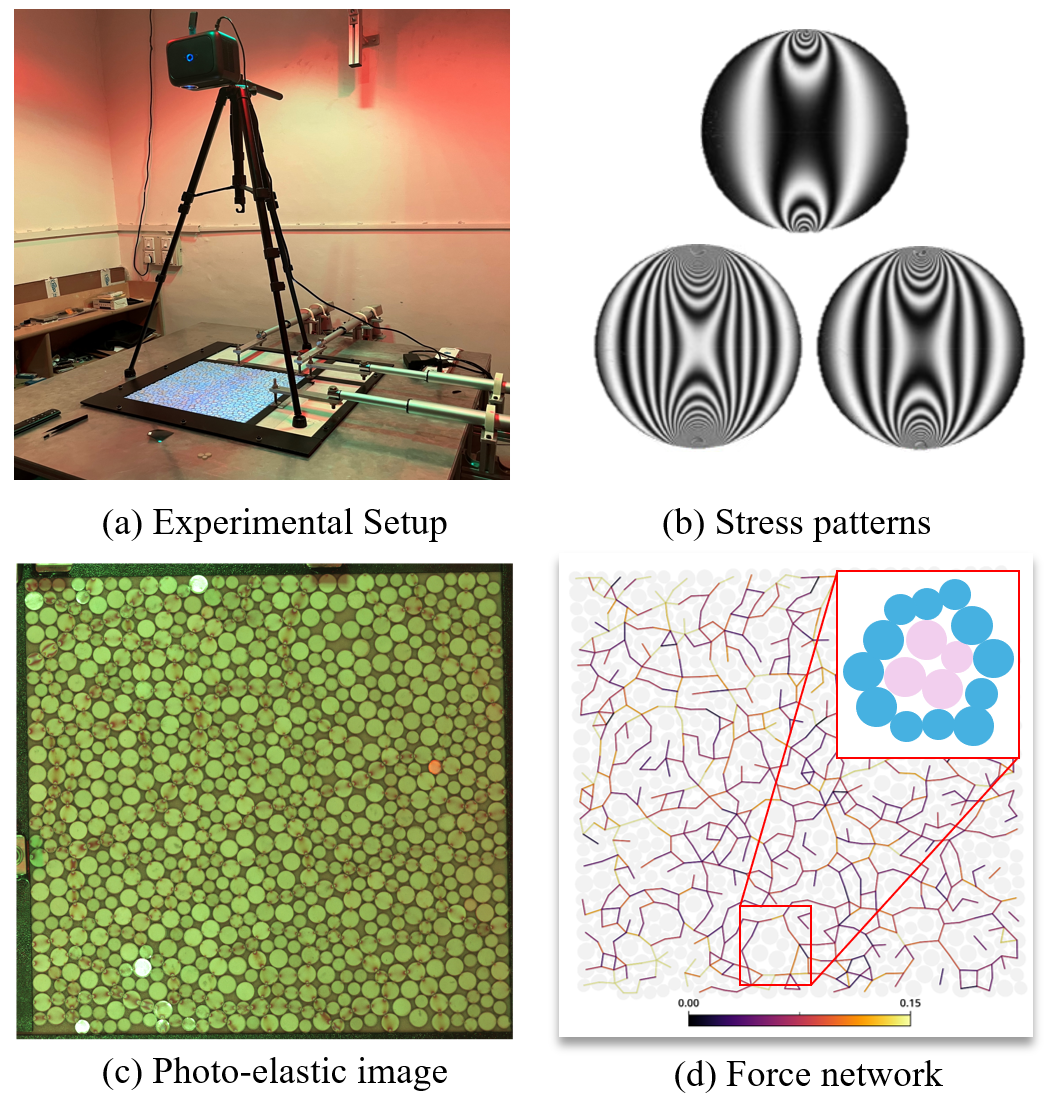}} 
    \caption{(a) The photoelastic experimental setup. (b) An example of the stress fringe pattern image on the disks observed through polarizing filters. It is adapted from~\cite{REN2020106263}. (c) The arrangement of the photoelastic disks at one of the time steps during the experiment, and, (d) the corresponding derived force network, \new{highlighting one cycle (blue disks) by a red square; the pink disks are the enclosed non-participating particles so-called rattlers}.  }
    \label{fig:data}
\end{figure}
\new{Granular materials consist of densely packed discrete particles, such as sand, gravel, or powder. Although each particle is solid, its collective behavior is complex and can resemble that of solids, fluids, or gases with different characteristics. These materials are essential in industrial applications, and understanding their behavior is crucial for optimizing processes and preventing issues like clogging, flow blockages,  and structural failure. When subjected to disturbances like shearing, vibration, or tapping, granular materials exhibit complex behavior due to uneven distribution of the forces propagating through the system. Unlike solids, liquids, or gases, the macroscopic properties of granular materials (e.g., friction, elasticity, plasticity) cannot be directly derived from microscopic properties (e.g., particle positions, velocities, inter-particle forces). To address this, a multi-scale approach has been developed, introducing a mesoscale level between the microscopic and macroscopic scales. The structures at these scales are clusters of particles, force chains (load-bearing connected particles), and force loops, referred to as cycles. These structures serve as the basis for averaging local properties to derive macroscopic properties. This work focuses specifically on force loops as a mesoscale structure.
\paragraph{Experimental setup and data.}
To better understand granular materials and how stress propagates through them, photoelastic experiments in two-dimensional systems are performed to quantify the contact forces between constituent particles~\cite{majmudar2005contact}. In these experiments, a collection of photoelastic disks is arranged in a tray~\cref{fig:data} and subjected to various loading conditions. The refractive index of the photoelastic disks changes in response to external stress, altering the polarization of light passing through the material and producing fringe patterns that indicate the paths of force transmission. These force pathways are directly visualized and captured by a high-resolution camera. The contact network, along with the magnitudes of forces at the contact points, is then approximated through complex post-processing steps, including fringe pattern analysis, image processing, and optimization problem-solving. For more details, we refer the reader to~\cite{daniels2017photoelastic}.}

\new{In our experiment, loads are applied to the system by displacing the left and top wall of the tray inwards followed by reducing the load by moving walls outward. The experiment keeps track of the positions of the disks and the fringe pattern on the disks~\cref{fig:data}(b) imaged under polarized light. They show the stress propagating in the system. After postprocessing, a series of force networks one for each loading condition are inferred from these patterns. We obtained the dataset as a weighted graph~\cref{fig:data}(d), where the nodes represent the disks' centers and the edges represent contacts between disks, weighted by the force value.
}

\section{Related Work and background}
\label{sesc:related}
In the following section, we first summarize some key work on the analysis of force networks in granular materials and then introduce general tracking frameworks that our research builds upon.

\paragraph{Cycles evolution in force network.}
In the granular material literature, cycles within the contact network are referred to as loops or mesoloops. In this paper, we use the term cycles for consistency with the topological framework. The evolution of cycles in the force or contact network has been studied to identify different states of the system under external influences~\cite{arevalo2010topology, arevalo2013contact}.
Tordesillas \etal~\cite{tordesillas2010force, tordesillas2012transition, tordesillas2014force} investigated the co-evolution of cycles and force chains (collections of grains interacting through relatively large forces) in the force network under various loading conditions. They observed that the number of 3-cycles (cycles comprised of three edges) and 4-cycles (cycles comprised of four edges) increases as the system transitions to a solid-like state.
While previous approaches based on cycle structures offer insight into the force network, the analysis is limited to fixed scales, leaving the question of finding appropriate scales open.
In contrast, we utilize an efficient method that extracts the entire hierarchy of these cycles allowing us to track and study the evolution of the entire hierarchy.

Also, a few algebraic topology approaches have been applied to force networks to identify different states in granular materials under external stress. 
Kondic \etal~\cite{kondic2012topology} were among the first to apply these methods for the analysis of granular materials. 
They investigated the evolution of force networks in slowly compressing granular materials as they undergo jamming, primarily focusing on the zeroth and first Betti numbers ($\beta_0, \beta_1$), which correspond to force chains and cycles in the force network, respectively, at distinct force threshold values. 
Additionally, they demonstrated how Betti numbers $\beta_0$ and $\beta_1$ can differentiate between frictional and frictionless systems. 
In subsequent work, they employed persistent homology to study the evolution of $\mathcal{H}_0$ and $\mathcal{H}_1$ in various system configurations while discussing features in the persistence diagram~\cite{kramar2013persistence}. 
The key concept of their filtration approach is presented in~\cite{kramar2014quantifying}, where they identify 3-cycles in their proposed chain complex by detecting triangles and cycles with more than three edges, referred to as defects. 
They further compared different states within the same system or across systems using bottleneck and Wasserstein distances between persistence diagrams~\cite{kramar2014evolution}. They also applied persistent homology to explore cycles and clusters in compressed tapped particulate systems~\cite{kondic2016structure, pugnaloni2016structure}.

More recently, Mei \etal~\cite{mei2023modeling} studied the transition of granular material from a solid-like state to a liquid state by looking at the $\mathcal{H}_0$ and $\mathcal{H}_1$ homology groups.
For more details, we refer the reader to an in-depth review of network analysis for granular materials~\cite{papadopoulos2018network}, where the authors reviewed the literature for network-based and algebraic topology-based methods for the study of granular materials, including the role of cycles or clusters in force network.

The commonality among these methods lies in their analysis based on counting cycles ($\beta_1$) in the force network as the system undergoes external loading. 
This counting approach has proven useful in quantifying force networks. However, these methods do not consider structural properties, such as anisotropy, loop elongation, or force tensor, of the loop structures within the force network, and fail to explain how these properties evolve at different scales and times. 
 This limitation arises because the representatives of homology groups are not unique, making it impossible to accurately compute and track the true geometric structures.
In contrast, our proposed method provides a unified framework that formalizes the extraction of desired representations of the cycles and summarizes their entire evolution at multiple scales using tracking graphs.

\paragraph{Feature tracking.}
Feature tracking is a common task in scalar field visualization, and topological data analysis methods are often used in the field. Generally, the temporal dimension is managed in one of two different ways \cite{Post2003}: features are extracted in each time step and then matched according to some criteria, or the temporal dimension is handled as an extra spatial dimension, thus the tracking becomes a spatio-temporal feature extraction \cite{Weber2011}. The two most common feature descriptors are features based on a point or a set of points, such as critical points, or spatial regions in the dataset.

Critical points have been used in vector field visualization to track vortices \cite{Weinkauf2011, Garth2004b}, and Reininghaus et al. \cite{Reininghaus2011} presented a combinatorial method for tracking critical points in a vector field using derived scalar fields to generate feature tracks. This method has been further extended to track hierarchical sets of critical points \cite{Engelke2020, Nilsson2022Cyclone}. Recent work in the field demonstrates that critical points can also be tracked by deriving similarity measures between the merge trees of subsequent time steps~\cite{Sridharamurthy2020, Pont2022, Wetzels2022a, Wetzels2023}. Critical points are contained within sub-/super levelsets, which Nilsson et al. \cite{Nilsson2023} employ to track extrema based on the overlap of the respective ascending/descending manifolds of the Morse-Smale complex. 

Spatial regions are traditionally defined by isocontours~\cite{lorensen1987marching} or sub-/super-/level sets based on topological descriptors~\cite{bremer2009analyzing, Sohn2006}. Early work by Samtaney et al.~\cite{Samtaney1994} summarizes tracking regions in a data set based on thresholds in a scalar field, augmented by geometric attributes such as boundary region, volume, and center of mass. They also employ the spatial overlap for tracking the features across time, which is utilized in a variety of applications. Silver and Wang~\cite{silver1998tracking} extended the work of \cite{Samtaney1994} to also work for unstructured datasets. Nested tracking graphs were introduced by Lukasczyk et al.~\cite{lukasczyk2017nested} to track nested isocontours, i.e. hierarchical features, using spatial overlap. Improving the layout of nested tracking graphs has also been a focus of recent work~\cite{Kopp2019, Dobler2024}. Spatial overlap has been used in conjunction with merge trees, for instance by Saikia and Weinkauf~~\cite{Saikia2017} to track subtrees in the merge tree where the overlap is used as the weight of the edges in a tracking graph. Widanagamaachchi et al.~\cite{Widanagamaachchi2012} also define feature tracks based on the spatial overlap of regions represented on a merge tree, but in contrast their method yields a tracking graph for a specific isovalue. Both methods yield a hierarchy by the use of the merge tree. However, the above approaches deal with datasets that have sparse features: they cover a small percent of the domain. 

A work with a similar goal to ours is the work of Schnorr et al.~\cite{Schnorr2020}, in which space-filling features are tracked and the entire region in a combustion simulation is divided into a set of features, establishing the correspondence based on spatial overlap. The resulting tracking graph has a high edge-to-node ratio, as a small perturbation in the field will yield overlap with multiple regions in the next time step. Feature tracks are established using a global optimization method to remove edges with lower overlap weights, however the procedure is computationally expensive. 

In contrast to the presented approaches, that track components~(corresponding to $\mathcal{H}_0$ homology groups), we track the cycles of a force network graph, which we can track by viewing the cycles as connected components of the dual complex, see~\cref{sec:extraction}. While we have space-filling features like Schnorr et al.~\cite{Schnorr2020}, we differentiate ourselves by tracking a nested hierarchy and using simple thresholding to refine the tracking results, which is faster than using global optimization.

\section{Multi-scale Cycle Extraction}
\label{sec:extraction}

A \emph{force network} can be represented as a weighted planar straight-line graph $G=(V, E)$ where the set of vertices corresponds to the centers of disks (or, in general, \emph{particles}) in a granular material system and the edges connect adjacent disks with non-zero force. The weight of an edge corresponds to the force acting between the incident disks. Note that there can be vertices in $V$ with no incident edges in $G$ denoting disks with no force acting on them, also called \emph{rattlers} in the domain of granular material science.

This section briefly summarizes the persistence homology-based method for the multi-scale extraction of cycles within force networks~\cite{rasheed2023multi}. The force network can be viewed as a one-dimensional simplicial complex $F$ with nodes (disk centers) being its 0-simplices while the edges (contacts) form its 1-simplices. Given that the force network is embedded in a 2D space, the focus lies only on the zeroth ($\mathcal{H}_0$) and first order ($\mathcal{H}_1$) homology groups. In a broad sense, these homology groups representing connected components and loops can be associated with mesostructures in granular materials, such as \emph{force loops}. These mesostructures are known to contribute to force distribution and stability in granular materials. 

To define filtration on $F$, 
we first need to assign a weight to each simplex in $F$. Let $\omega$ be the maximum weight of an edge in the force network and $\varepsilon$ is a negligibly small number. All the vertices are assigned weights equal to $\omega+\varepsilon$, while the edges get the force values as their weights.
Let $F_\alpha$ be the subcomplex of $F$ consisting of simplices having a weight greater than or equal to $\alpha$. 
Then varying $\alpha$ from 0 to $\omega + \varepsilon$ results in the following filtration, a finite sequence of nested simplicial complexes:
\begin{equation}
\label{eqn:FN_filtration}
    F=F_{0} \supseteq F_{\alpha_1} \supseteq F_{\alpha_2}  \supseteq ... \supseteq F_{\alpha_k}=F_{\omega+\varepsilon} \supseteq \emptyset
\end{equation}
where $\alpha_i$ represents the force threshold with $\alpha_i<\alpha_{i+1}$ and $F_{\alpha_i}$ denotes the force network with edges above weight $\alpha_i$, or to draw a parallel to scalar fields, a `super-level set' of force network. 

Typical persistent homology-based analysis of $F$ can be used to determine the $\mathcal{H}_1$ homology group, its generators, and their importance based on persistence. However, our interest lies in the extraction of a more descriptive understanding of the mesoscale features within the force network, for which we need a consistent set of generators (cycles) that partition the granular material system. 
Therefore, we employ a dual filtration approach, introduced in \cite{rasheed2023multi}, that exploits Alexander's duality theorem and implicitly extracts the relevant representatives of $\mathcal{H}_1$ homology group within the force network. 
Alexander's duality theorem establishes a correspondence between the $\mathcal{H}_1$ homology group of $F$ and the $\mathcal{H}_0$ homology group of the dual $S^2-F$, where $S^2$ is a 2D sphere. 
That is, instead of directly computing cycles in the force network $F$, connected components can be extracted in $S^2-F$ and be used as representatives of the cycles in $F$. 
The advantage of computing homology groups through dual filtration is that we obtain a concise representation of the homology generators that remain consistent across filtration levels. In the following, we offer a brief overview to get intuition about the method. For more details, we refer the reader to~\cite{rasheed2023multi}.

First, the \emph{constrained Delaunay triangulation} $D$ of the vertices $V$ is computed ensuring that the edges $E$ of the force network are part of the triangulation $D$. The edges in the $D-F$ are assigned zero weight.
Similarly, the triangles in $D-F$ are also assigned a zero weight.
Now consider the dual graph $\hat{G}$ of $D$ where the triangles in $D$ form the vertices of $\hat{G}$ while edges connect the adjacent triangles. 
With the weights as described, the following filtration is induced on $\hat{G}$:
\begin{equation}
\label{eqn:dual_filtration}
    \emptyset \subseteq \hat{G}_0 \subseteq \hat{G}_{\alpha_1} \subseteq \hat{G}_{\alpha_2}  \subseteq ... \subseteq \hat{G}_{\alpha_k}=\hat{G}_{\omega+\varepsilon} = \hat{D}
\end{equation}
Compare the two filtrations in~\ref{eqn:FN_filtration} and~\ref{eqn:dual_filtration}  and observe the duality between the force network filtration and the induced filtration on the dual graph $\hat{G}$. At $\alpha=0$, the set of all connected components in $\hat{G}_0$ partitions the underlying domain $D-F_0$, where the boundaries of each partition correspond to the cycles in the force network $F_0$, and the enclosed vertices represent non-participating particles, also known as rattlers. This representation of cycles ensures consistency in cycle generators across multiple scales and facilitates the direct computation of various mesostructure characteristics. 

Consider the alternate perspective of the evolution of cycles as the filtration parameter value $\alpha$ is varied. All the cycles in the force network are born at the filtration parameter value $\alpha = 0$ and as the value of $\alpha$ increases, the connected components of the dual graph $\hat{G}$ start merging and, finally, at maximum $\omega+\varepsilon$ end up with a single connected component. This development of the cycles and cycle-bounded segments can be represented by a tree that is similar to \emph{merge tree} in the context of scalar fields. 
Each node represents a cycle-bounded segment of the domain, i.e., a set of triangles belonging to the same connected component. The links between nodes in this tree indicate parent-child relationships, demonstrating the hierarchy or nesting of cycle-bounded segments. We refer to this tree as the \emph{cycle hierarchy} of the force network $F$ and in the rest of the paper, we will use the notation $\mathcal{C}=(N, L_h)$ to denote the cycle hierarchy of a given force network where $N$ is the set of nodes in the tree $\mathcal{C}$ and $L_h$ are the edges or hierarchy links.

\section{Multi-scale Cycle Tracking}
\label{tracking}

\begin{algorithm}[!ht]
    \caption{Compute Tracking Graph }\label{alg:three1}
    \textbf{Input: } A time series of force networks $F^1, F^2, \ldots, F^T$\\
    \textbf{Input: } $Levels = (\alpha_1 \leq \alpha_2 \leq \ldots \leq \alpha_k\}$\\
    \textbf{Output: $\mathcal{G}=(\mathcal{N}, \mathcal{L}_h, \mathcal{L}_t)$} \\
    $Nodes \leftarrow \emptyset$\\
    $HierachyLinks \leftarrow \emptyset$\\
    $TemporalLinks \leftarrow \emptyset$\\
    \For{$t\leftarrow 1$ \KwTo $T$}{
        $D^t \leftarrow ConstrainedDelaunayTriangulation(F^t)$\\ 
        $\mathcal{C}^t \leftarrow GenerateCycleHierachy(F^t, D^t)$\\
        ($N^t, L^t_h) \leftarrow RestrictedCycleHierachy(\mathcal{C}^t, Levels)$\\
        $Nodes \leftarrow Nodes \cup N^t$\\
        $HierachyLinks \leftarrow HierachyLinks \cup L^t_h$\\
    }
    \For{$t\leftarrow 1$ \KwTo $T-1$}{
        \CommentSty{\textit{// Compute overlap of the triangles}}\\
        $I \leftarrow size(D^t); J \leftarrow size(D^{t+1})$\\
        $M_{I \times J} \leftarrow \{0\}_{I \times J}$\\
        \For{$i\leftarrow 1$ \KwTo $I$}{
                \For{$j\leftarrow 1$ \KwTo $J$}{
                     $\triangle_i \leftarrow D^t[i]$\\
                     $\triangle_j \leftarrow D^{t+1}[j]$\\
                    $M[i, j]\leftarrow computeOverlap(\triangle_i, \triangle_j)$
                    }
                }
        
        \BlankLine
        \CommentSty{\textit{// Compute overlap of leaf nodes}}\\
        $C^t_{\alpha_1} \leftarrow GetCyclesAtLevel(C^t, \alpha_1)$\\
        $C^{t+1}_{\alpha_1} \leftarrow GetCyclesAtLevel(C^{t+1}, \alpha_1)$\\
        $\mathcal{L}_1 \leftarrow ComputeRegionOverlap(M, C^t_{\alpha_1}, C^{t+1}_{\alpha_1}, \rho)$\\
        $TemporalLinks \leftarrow TemporalLinks \cup \mathcal{L}_1$
        \BlankLine
        \CommentSty{\textit{// Compute Links for higher levels}}\\
        \For{$l\leftarrow 2$ \KwTo $k$}{
            $\mathcal{L}_l\leftarrow \emptyset$\\
            \For{$e_{l-1} \in \mathcal{L}_{l-1}$}{
                $e_{l} \leftarrow (Parent(e_{l-1}.n_1), Parent(e_{l-1}.n_2))$\\
                $\mathcal{L}_l \leftarrow \mathcal{L}_l \cup \{e_l\}$\\
                $UpdateOverlap(e_l)$\\
                
            }
            $TemporalLinks \leftarrow TemporalLinks \cup \mathcal{L}_l$
        } 
        }
    \Return $\mathcal{G}=(Nodes, HierarchyLinks, TemporalLinks)$
    \BlankLine
    \SetKwProg{Function}{Function}{:}{end}
    \Function{ComputeRegionOverlap $(M, C_1, C_2, \rho)$}{
    $E \leftarrow \emptyset$\\
    \For{$k\leftarrow 0$ \KwTo $size(C_1)$}{
        \For{$l\leftarrow0$ \KwTo $size(C_2)$}{
            \BlankLine
            $Overlap(C_1[k], C_2[l]) \leftarrow \sum_{\triangle_i\in C_1[k]} \sum_{\triangle_j\in C_2[l]} M_{ij}$
             \BlankLine
            $\Omega \leftarrow \dfrac{Overlap(C_1[k], C_2[l])}{\min(Area(C_1[k])), Area(C_2[l]))}$\\

            \If{$\Omega>\rho$}{
                $E\cup\{(C_1[k].label, C_2[l].label)\}$
                }                      
            }
                
        }
    \Return $E$
    }
                      
    \end{algorithm}

\begin{figure*}
    \centering
    \includegraphics[width=.8\textwidth]{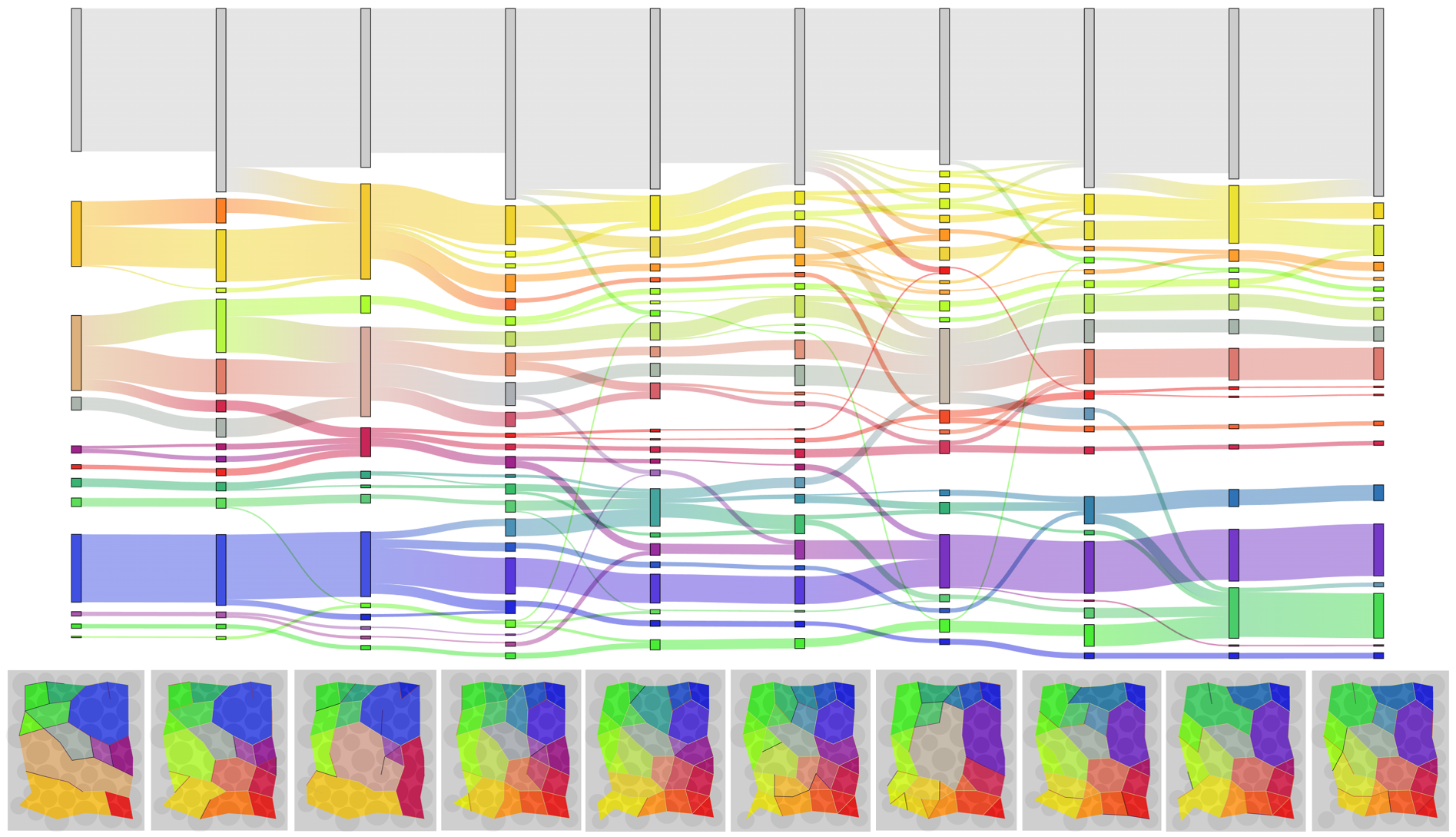}
    \caption{Dataset \textit{0} - Sankey diagram visualization for one level. The ten columns represent different time steps (loading/unloading). The nodes in a single column indicate the respective cycles/segments. 
    The width of the links represents the amount of overlap. The nodes and links are colored according to the spatial 2D color map, emphasizing the location of the cycles and helping to show how nearby cycles are developing over time.}
        \label{fig:s3_sankey}
\end{figure*}

In this section, we describe our proposed method to track the evolution of cycles in force networks in time-dependent contexts. The method exploits the definition of cycles through the dual approach as described in the previous section. The resulting partitioning of the domain into non-overlapping cycle-bounded segments is crucial for tracking changes over time. The method is inspired by the concept of nested tracking graphs~\cite{lukasczyk2017nested} that have been used to track and visualize regions in scalar fields at multiple scales over time. However, there are some key challenges that need adaptation in order to apply these ideas to our context.

Given a time series of force networks $\{F^t \mid t=1, 2, \ldots, T\}$ representing an experiment or simulation of propagation of forces and stresses in a granular material system. Let $D^t$ and $\mathcal{C}^t$ denote the Delaunay triangulation and cycle hierarchy at time step $t$ computed from the method described in~\cref{sec:extraction}. Moreover, an input parameter $k > 1$ is provided that is used to restrict the cycle hierarchy $\mathcal{C}$ to $k$ levels at filtration thresholds $A = (\alpha_1 \leq \alpha_2 \leq \ldots \leq \alpha_k)$. Note the exact values of $\alpha_l, 1\leq l \leq k$, can be kept the same for all time steps or they can vary based on time. 
\new{In this work, we choose two levels corresponding to a fine level ($\alpha_1=0$) and a coarse level with $\alpha$ set equal to the median of the contact force values in the force network $F^t$.}
A simpler restricted cycle hierarchy tree $\mathcal{C}(A)$ can constructed from the complete cycle hierarchy $\mathcal{C}$ that represents cycles extracted only at filtration thresholds $\alpha_l\in A$ and their nesting hierarchy.

To explain the concept of temporal tracking of cycles, let's focus on one level $l$ in cycle hierarchy defined by threshold $\alpha_l$. 
To construct a tracking graph at this level $l$, we need to define the correspondence between nodes at level $l$ across adjacent time points.
This correspondence can be modeled as a bipartite graph. 
Let $N^t$ and $N^{t+1}$ be the set of nodes at time points $t$  and $t+1$ respectively. A link connecting a node $n \in N^t$ to a node $m\in N^{t+1}$ is added if there is some spatial overlap between the corresponding cycle-bounded segments of the two nodes. 
We use the \emph{overlap coefficient}, $\Omega(n, m)=|n\cap m|/\min(|n|, |m|)$ to define the importance of a specific temporal link. 
This coefficient is particularly useful because it highlights the importance of smaller regions, such as a single triangle, which contributes to the stability of the granular ensemble and whose tracking is essential for understanding stability over time. 
The nodes having an overlap coefficient greater than a user-specified threshold $\rho$ are connected by temporal links.

To efficiently generate correspondence between cycle-bounded segments in consecutive time steps, we first compute the pairwise overlap of each triangle in $D^t$ with each triangle in $D^{t+1}$. This results in an overlap matrix $M_{I \times J}$, where $I$ and $J$ are the numbers of triangles in $D^{t}$ and $D^{t+1}$ respectively. The entry $M_{i,j}$ in the matrix stores the area of overlap between triangle $\triangle_i$ in $D^t$ and triangle $\triangle_j$ in $D^{t+1}$.
We then compute the tracking graph for the leaf nodes of $\mathcal{C}^{t}(A)$, i.e. the nodes at level $l=1$.
Using the overlap matrix $M$, we compute the region overlap between $m\in \mathcal{C}^t_{\alpha_1}$ and $n\in \mathcal{C}^{t+1}_{\alpha_1}$ based on the sum of overlaps of their respective triangles.
Formally, let $S(m)$ and $S(n)$ be the set of triangles associated with node $m$ and $n$, respectively. The $\emph{overlap coefficient}$ $\Omega(m, n)$ is defined as:
\begin{equation}
    \Omega(m, n) = \frac{\sum_{\triangle_i\in S(m)} \sum_{\triangle_j\in S(n)} M_{ij}}{\min(\sum_{\triangle_i\in S(m)}Area(\triangle_i), \sum_{\triangle_j\in S(n)}Area(\triangle_j))}
\end{equation}

A link $e=(m, n)$ is established between $m$ and $n$ nodes of two adjacent time points if the region overlap coefficient satisfies a given threshold $\Omega(n,m)>\rho$. The final set of temporal edges at the leaf level can be constructed by computing the links for all adjacent time steps. Formally, 
$\mathcal{L}_1 = \{(n,m) | \Omega(n,m)>\rho $ for $n\in \mathcal{C}^t_{\alpha_1}$ and $m\in \mathcal{C}^{t+1}_{\alpha_1}\}$ denotes the set of all temporal edges at the leaf hierarchy, i.e., level 1.

It is important to note that we only need to compute the temporal links for the leaf level. The temporal edges $\mathcal{L}_2, \mathcal{L}_3, \ldots, \mathcal{L}_k$ for higher levels $\alpha_2 \leq \alpha_3 \leq \ldots \alpha_k$ can be inferred from the level hierarchy information along with the temporal edges at the leaf level $\mathcal{L}_1$. Specifically, if a link is established between node $n\in \mathcal{C}^{t}_{\alpha_l}$ and $m\in \mathcal{C}^{t+1}_{\alpha_l}$, then we can infer the link $e'=(parent(n), parent(m))$, where $parent(n)\in \mathcal{C}^{t}_{\alpha_{l+1}}$ is the parent node of $n$. The overlap information and size of the nodes for each level can also be computed by aggregating the node size and overlap. Since the nodes correspond to a space partitioning, the size of the parent node is the sum of the size of all child nodes. Here, we would like to emphasize that this property does not hold for the nesting of sublevel sets in scalar fields for which the original nested tracking graph was proposed~\cite{lukasczyk2017nested}.

The final output is the nested tracking graph $\mathcal{G}=(\mathcal{N}, \mathcal{L}_h, \mathcal{L}_t)$ where $\mathcal{N}$ is the set of nodes, $\mathcal{L}_h$ is the set of all links connecting nodes in adjacent hierarchy levels, and $\mathcal{L}_t$ is the set of all temporal links. More specifically, for the given time series of restricted cycle hierarchy trees $\mathcal{C}^t(A)=(N^t,L_h^t)$, the sets  $\mathcal{N}$, $\mathcal{L}_h$ and $\mathcal{L}_t$ are $\mathcal{N} = \bigcup_{t=1}^T N^t$, $\mathcal{L}_h = \bigcup_{t=1}^T L^t_h$, and $\mathcal{L}_t = \bigcup_{l=1}^k \mathcal{L}_l$, respectively.

\paragraph{Implementation details.}
The input to our tracking algorithm is a series of force networks $F^t$ from which a series of Delaunay triangulations $D^t$ and the corresponding hierarchical partitioning $\mathcal{C}^t$ are computed. We assign a unique label to each node in every hierarchy tree $\mathcal{C}^t$. This labeling is crucial in laying out a tracking graph.
We constructed hierarchical partitioning using \emph{union–find} data structure, such that each node in the tree has the information of its parent. In addition, the leaf nodes also have the information of the respective cycle-bounded segments given in the form of a set of connected triangles.
The information on cycle-bounded segments of internal nodes can be determined by aggregating the leaf nodes. 

The details of the algorithm are provided in ~\autoref{alg:three1}. \new{The worst-case runtime complexity of this algorithm is quadratic in the number of disks (or particles) constituting the granular system. Let the number of disks be $n_d$. Then, the computation of Delaunay triangulation can be performed in $O(n_d\log n_d)$ time~\cite{aurenhammer2013voronoi}. Moreover, due to planarity, the number of triangles and edges in the resulting Delaunay triangulation has a linear upper bound i.e. $O(n_d)$. The most time-consuming step in this implementation is the computation of the overlap matrix $M$ which involves computing the overlap between triangles in consecutive time steps. In the worst case, this step is quadratic in the number of triangles, or $O(n_d^2)$. However, in practice, the number of overlapping triangles is closer to linear than quadratic due to the almost uniform distribution of disks in most granular systems. Grid-based spatial data structures can further accelerate the search for overlapping triangles}. In our implementation, for increased efficiency, we check if there is an overlap between the bounding boxes of two triangles before checking for a triangle-triangle intersection. In case of intersection, we compute the area of overlap using the open source library Shapely~\cite{shapely}.
\section{Visual Exploration of the Tracking Graph}
\label{sec:visualization}

We describe the visualization of the tracking graph within an interactive exploration framework, which includes the tracking graph, partitioning representation, and hierarchical structure using trees. All views are linked, allowing for selection and highlighting of tracks and segments, with various color schemes. Parameters for calculating the tracking graph can be adjusted interactively.

\subsection{Cycle tracking graph visualization}
The cycle tracking graph provides an overview of the evolution of cycles/segments over time, highlighting merge and split events. Internally, it is represented by nodes and links: nodes represent distinct segments at selected levels across all time steps, while links depict the cycle hierarchy and segment relationships between time steps. Nodes and links can include additional attributes like size, spatial location, or overlap.
In the visualization, nodes are arranged with time on the horizontal axis and a one-dimensional projection of domain partitioning on the vertical axis. The layout balances minimizing edge crossings with computational efficiency. We offer two layout options: the nested tracking graph~\cite{lukasczyk2017nested} and the Sankey diagram~\cite{Kennedy1898}.

\paragraph{Nested tracking graph.}
Consistent node labeling is crucial for drawing the tracking graph, as it forms the basis for the layout and coloring algorithm. Labels are typically initialized at the first time step and propagated over time. During merge and split events, it's necessary to identify which node retains the label and which receives a new one. This process is challenging when many events occur simultaneously, making it difficult to determine a valid successor.
To avoid inconsistent track coloring due to poor label choices, we propose a 2D colormap independent of labels, encoding the segments' spatial positions. Details on node and link color assignments are provided below.
\begin{figure}[htb]
    \centering
    \scalebox{1}[-1]{\includegraphics[width=\linewidth]{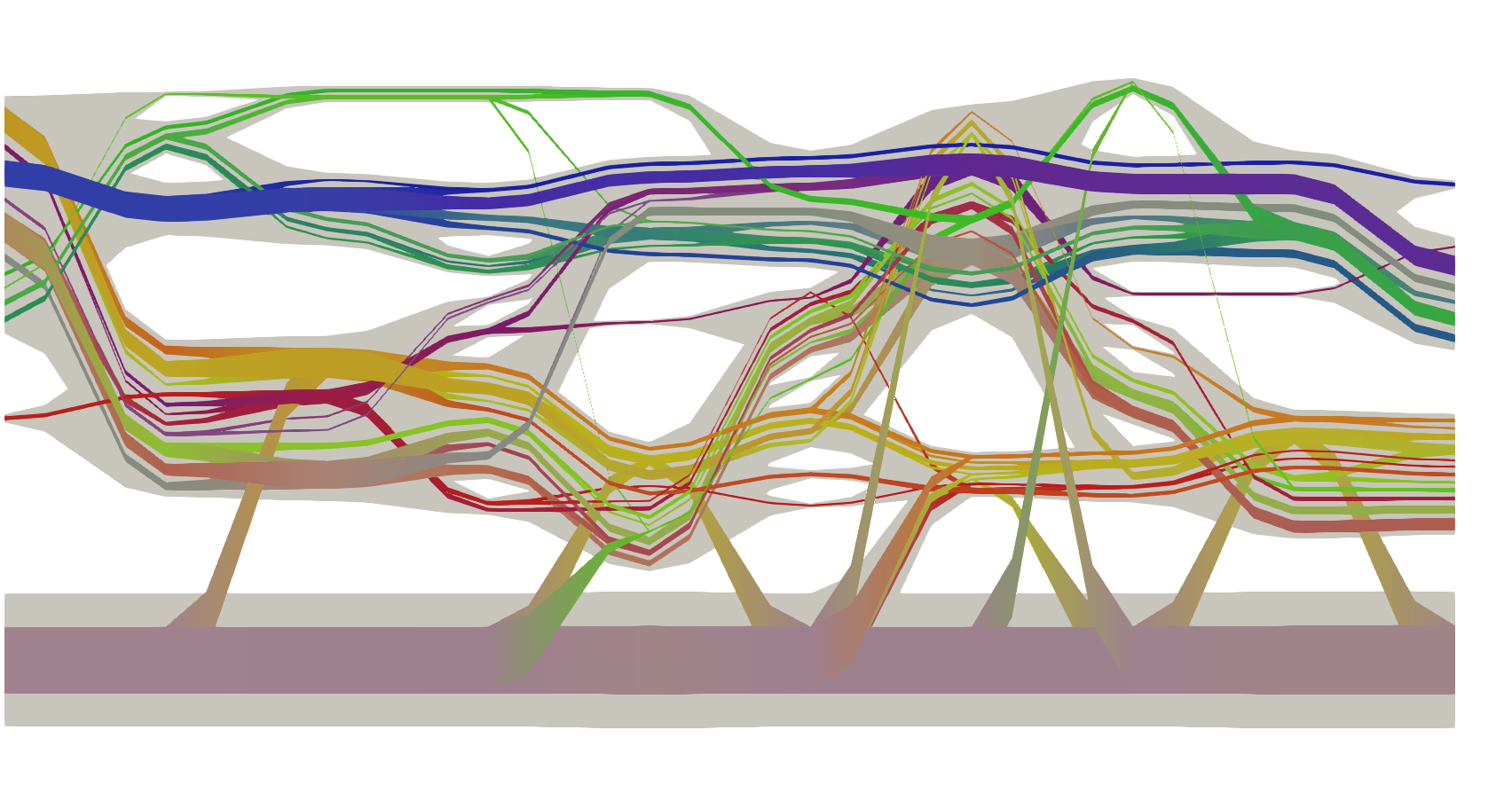}} 
    \caption{Nested Tracking Graph Visualization (Dataset \textit{0}): The graph displays ten time steps across two levels. Coarser-level tracking is shown in grey, while finer-level tracking uses a 2D colormap. Segments are connected if their overlap exceeds a threshold. As overlap values are computed separately for each level, the graph is not strictly nested. Link width reflects segment size between time steps. \new{To emphasize nesting for the partitioning, nodes at coarser levels are scaled by a small factor}.
    }
    \label{fig:s3_nested}
\end{figure}

\paragraph{Sankey diagram.}
Since the nested tracking graph doesn't focus on individual time step segmentation, we also provide a visualization using a Sankey diagram, adapted from D3~\cite{d3}. This effectively illustrates the overall development of partitioning over time at one level.
Like the nested tracking graph, the Sankey diagram positions nodes horizontally by time and vertically to minimize edge crossings. Nodes are displayed as thin rectangles, scaled by segment area, providing a one-dimensional projection of domain partitioning at each time point, highlighting size, connectivity, and spatial location. Parent segment coloring adds information about neighboring segments.
Links between nodes represent overlap, with link thickness proportional to the overlap value, indicating the connection strength. Node and link coloring can be independently determined to encode additional information (details see below).

\begin{figure}[htb]
    \centering
    {\includegraphics[width=\linewidth]{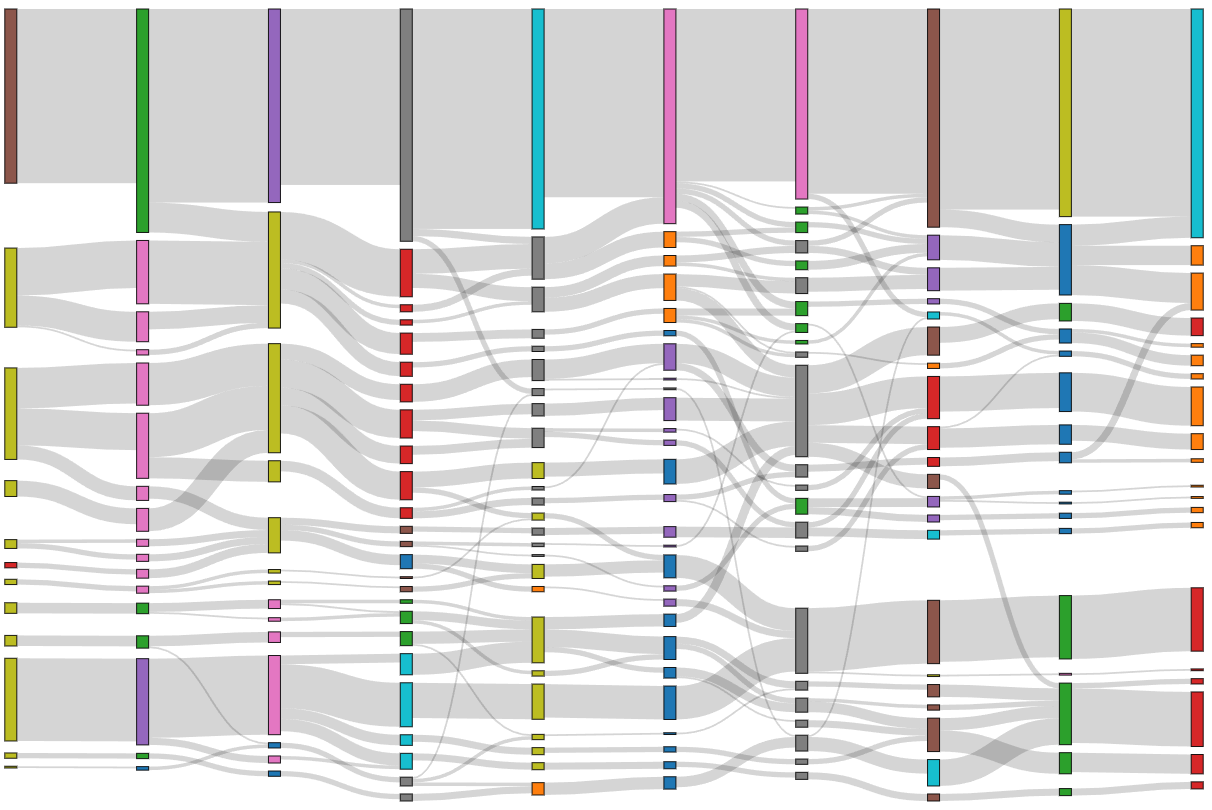}}
    \caption{Dataset \textit{0} - Sankey diagram visualization. The parameters for the visualization are similar to those in~\cref{fig:s3_sankey} but here the node colors refer to the label of the parent node in the hierarchy, emphasizing groups of segments that have the same parent.}
   
    \label{fig:small_parnt_color}
\end{figure}

\begin{figure}[!h]
    \centering
    {\includegraphics[width=.5\linewidth]{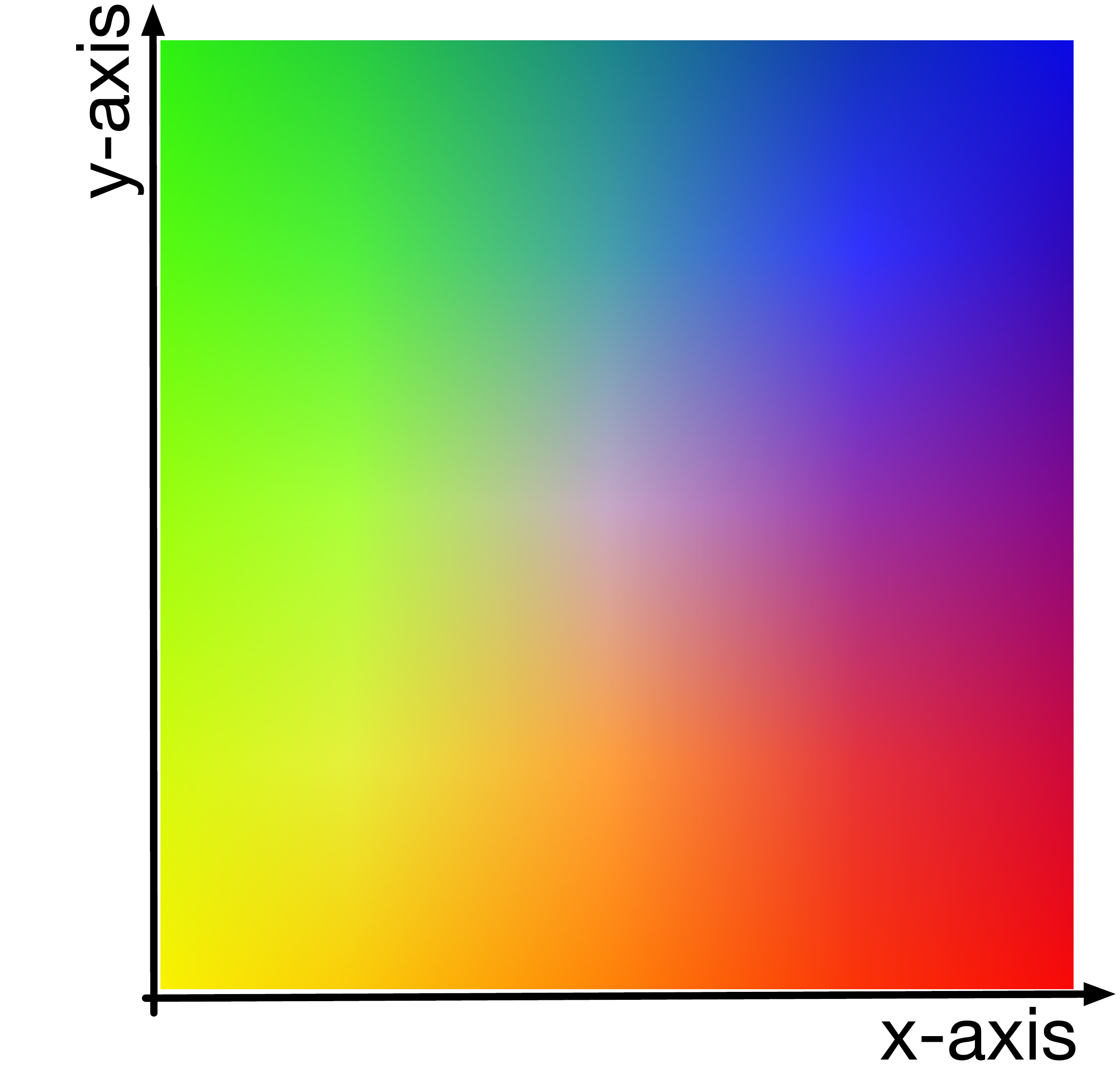}} 
    \caption{Two-dimensional color map used to encode the spatial location of the centroid of the segments.}
    \label{fig:colormap}
\end{figure}

\paragraph{Coloring of the tracking graph.}
\label{sec:color}
Coloring the tracking graph is essential for providing context and encoding additional attributes of nodes and links. We primarily use color to represent the spatial location of nodes and the nesting hierarchy. 
Defining a consistent coloring based on node or link labels is not trivial we propose a 2D colormap, see Fig.~\ref{fig:colormap}, that directly encodes the $(x,y)$ coordinates of the centroid of the segments. 
The colormap ensures that nodes representing nearby segments have similar colors, while those farther apart have distinct colors. Segments maintain consistent colors over time, as they usually move only slightly. Merging segments with similar sizes results in interpolated colors, while merging small segments into larger ones primarily retains the color of the larger segment.
This enhances the interpretability of the tracking graph and provides spatial context for each node's location in the data domain. We offer this colormap for both tracking graph layouts: the nested tracking graph in~\cref{fig:s3_nested} and the Sankey diagram in~\cref{fig:s3_sankey}.
To select an appropriate 2D colormap, we referred to the study by Steiger \etal~\cite{steiger2015explorative}. We chose the colormap (see~\cref{fig:colormap}) that best highlights distinct regions within our partitioning, enhancing the overall clarity of our visualization.

We also offer a coloring strategy to highlight node hierarchy at individual time points. Nodes with the same parent in the higher hierarchy are assigned the same color, making hierarchical relationships visible. This is especially useful for the Sankey diagram, see~\cref{fig:small_parnt_color}, while the nested tracking graph inherently shows hierarchy through its structure. Additionally, any available attribute can be used for node and link coloring if needed.

\subsection{Interaction and exploration}

In addition to the tracking graph, we include spatial representations of the domain for selected time steps, showing disk arrangements and the force network with edges colored by force values ~\cref{fig:data}(a,b). An alternative view displays cycle-bounded segments, which can be colored randomly or with the 2D colormap used in the tracking graph. The partition hierarchy is visualized as a tree, with nodes colored to match those in the tracking graph and cycle-bounded regions. Users can interactively select the filtration threshold to update the rendering based on the tree structure.
This visual analysis framework supports interactive exploration, allowing users to switch color schemes (static, hierarchical, or 2D colormap) via a drop-down menu. Hovering over a node reveals a tooltip with details and highlights all related links, showing the node's development over time (, see~\cref{fig:colormap_highlight}. Users can also select filtration values by percentile range and explore additional interaction options and statistical plots as described in~\cite{rasheed2023multi}. The framework is implemented as a prototype in Inviwo~\cite{Jonsson2020b}.

\section{Case study - Photoelastic Disks Experiment}

\begin{figure}
    \centering
    {\includegraphics[width=.8\linewidth]{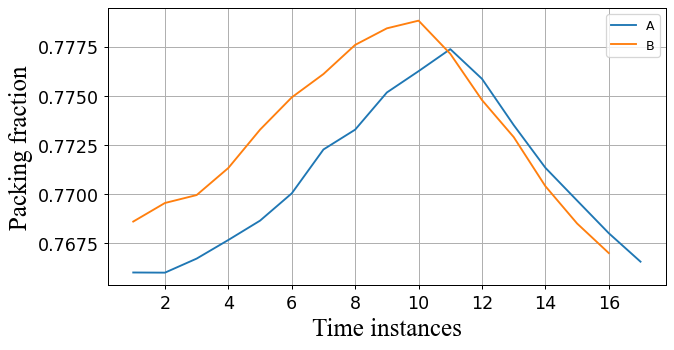}}
    \caption{Record of packing fraction of experiments $A\&B$ during loading and unloading process for experiment A and B respectively.}
    \label{fig:packingfrac}
\end{figure}
The datasets utilized in this study were generated through photoelastic disk experiments conducted by collaborating domain experts. We demonstrate our method on the dataset result from three experiments: \textit{0}, \textit{A} and \textit{B}. 

\begin{figure*}
    \centering
    {\includegraphics[width=\textwidth]{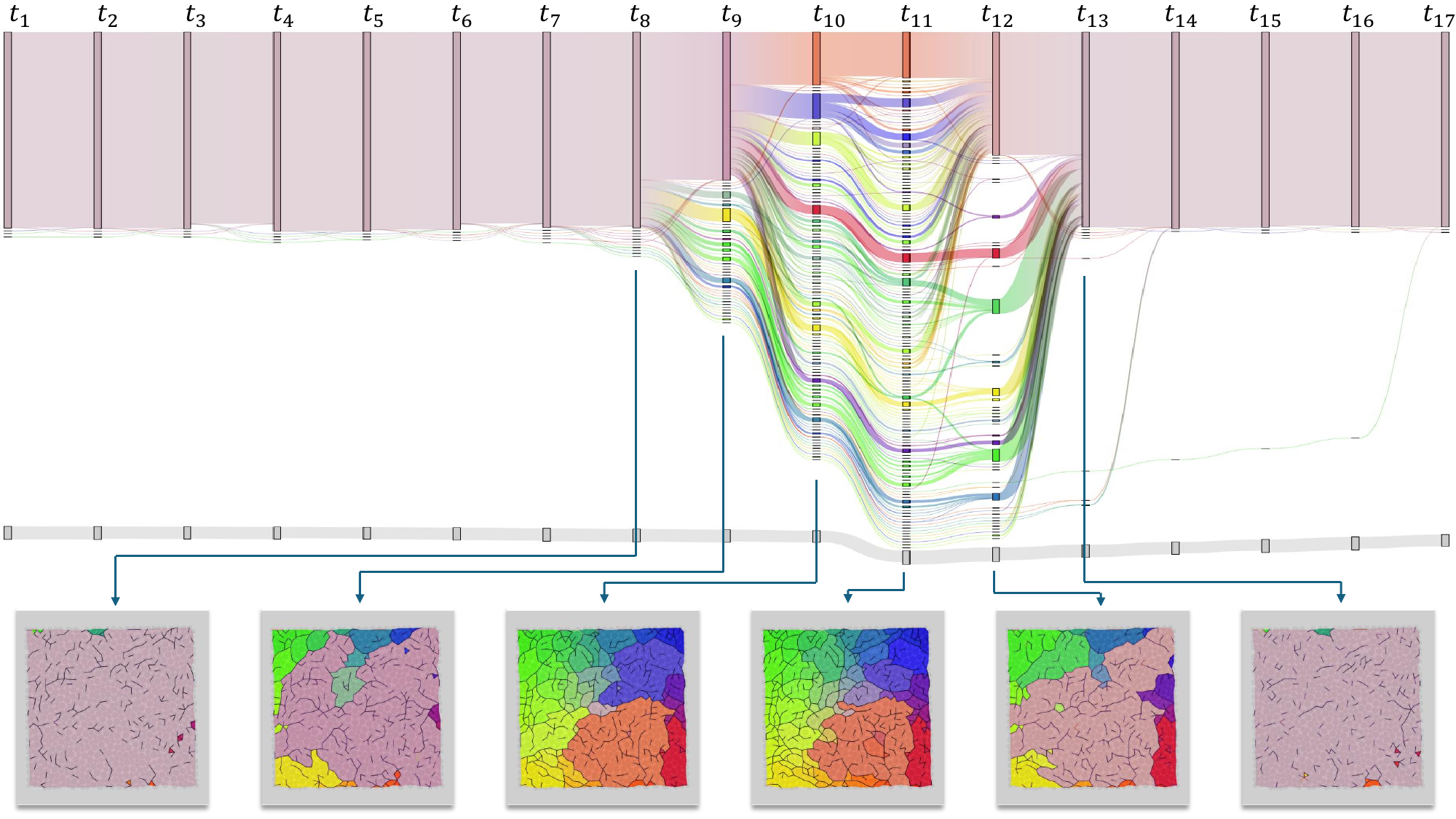}} 
    \caption{Experiment A: (top) Tracking graph visualization of a granular system's transition to jamming followed by unjamming. 
    (Bottom) Rendering of the domain partitioning into cycle-bounded segments. From time point $t_8$ the system transits into the jamming phase, which can be observed by the splitting of the domain into many segments.  }
    \label{fig:s3}
\end{figure*}

Experiment \textit{0} is a small data set that is a small section of a larger experiment available for 10 time steps. It is mainly used for illustrative purposes~\cref{fig:s3_sankey,fig:s3_nested,fig:small_parnt_color}. 
Experiment \textit{A} and \textit{B} are examples from a large ensemble of experiments with data for 16 and 17 time steps respectively. 
Tracking graphs for experiments $A$ and $B$, together with their partitioning snapshots are shown in~\cref{fig:s3} and ~\cref{fig:teaser}. 

Our collaborators are interested in understanding the differences or similarities in the dynamics of the force network when the system undergoes a jamming state while applying similar external loads. 
The transition of the jamming state is of special interest as it can be understood as a kind of phase transition where the characteristics of the granular material change substantially.
Typically, the jammed and unjammed states are identified by a measure called packing fraction, the ratio of the area occupied by the disks, and the total area of the system. 
A higher packing fraction indicates the jamming phase. The development of the packing fraction of experiments $A\&B$ is presented in~\cref{fig:packingfrac}.
In experiment $A$, the system transits into the jamming state at $t_8$ and back to the unjammed state at $t_{13}$ while for experiment $B$, the system is in the jamming phase from $t_6$ to $t_{12}$.
These observations were validated by the domain collaborators.
In addition to identifying the jammed and unjammed states, we further investigate the following questions using the tracking graph:
(1) Spatial analysis: What are the most jammed regions in the system? (2) Temporal analysis: How do specific segments develop under loading? What is the history of individual disks?
The proposed explorative tracking environment is well-suited for investigating these scenarios.

\paragraph{(1) Local jamming}
Our first study focuses on analyzing the local development of jamming. The hypothesis is that in jammed regions, there is strong force propagation with fewer non-participating disks, which is evident through an increase in the number of smaller cycle-bounded segments. This behavior is clearly illustrated in the tracking graph for dataset \textit{B}, as shown in~\cref{fig:teaser}. In this example, the transition to the jamming phase begins around $t_5$ and $t_6$. During this period, green-colored links first appear in the tracking graph, emphasizing the splitting of segments in the upper left corner of the system. The 2D colormap helps associate these behaviors with their respective areas in the dataset. This observation aligns with the fact that the load is applied on the left and top walls, resulting in force propagation from left to right and top to bottom. At time $t_7$, there are still two regions (yellow and purple) that remain largely unjammed. By $t_8$, jamming is visible throughout the entire dataset. Comparing these results with the second dataset \textit{A} in~\cref{fig:s3}, we observe different behavior: some regions, such as the orange region, remain unjammed throughout the entire experiment.

\paragraph{(2) Temporal evolution and history of rattlers}
The focus on so-called rattlers, the disks contained inside a cycle, provides a complementary view of this behavior. Rattlers are disks that either do not or weakly interact with other disks. These rattlers do not directly contribute to force propagation in the system.
In the tracking graph, large segments having a large width suggest the existence of the rattlers.
In both experiments, we see most of the rattlers on the right side of the domain where the size of the loops is large. However, for experiment \textit{A}, we note that the red, yellow, and orange branches in the tracking graph do not split much, indicating that the rattlers within these cycles stay rattlers during the entire experiment.

\begin{figure}
    \centering
    {\includegraphics[width=\linewidth]{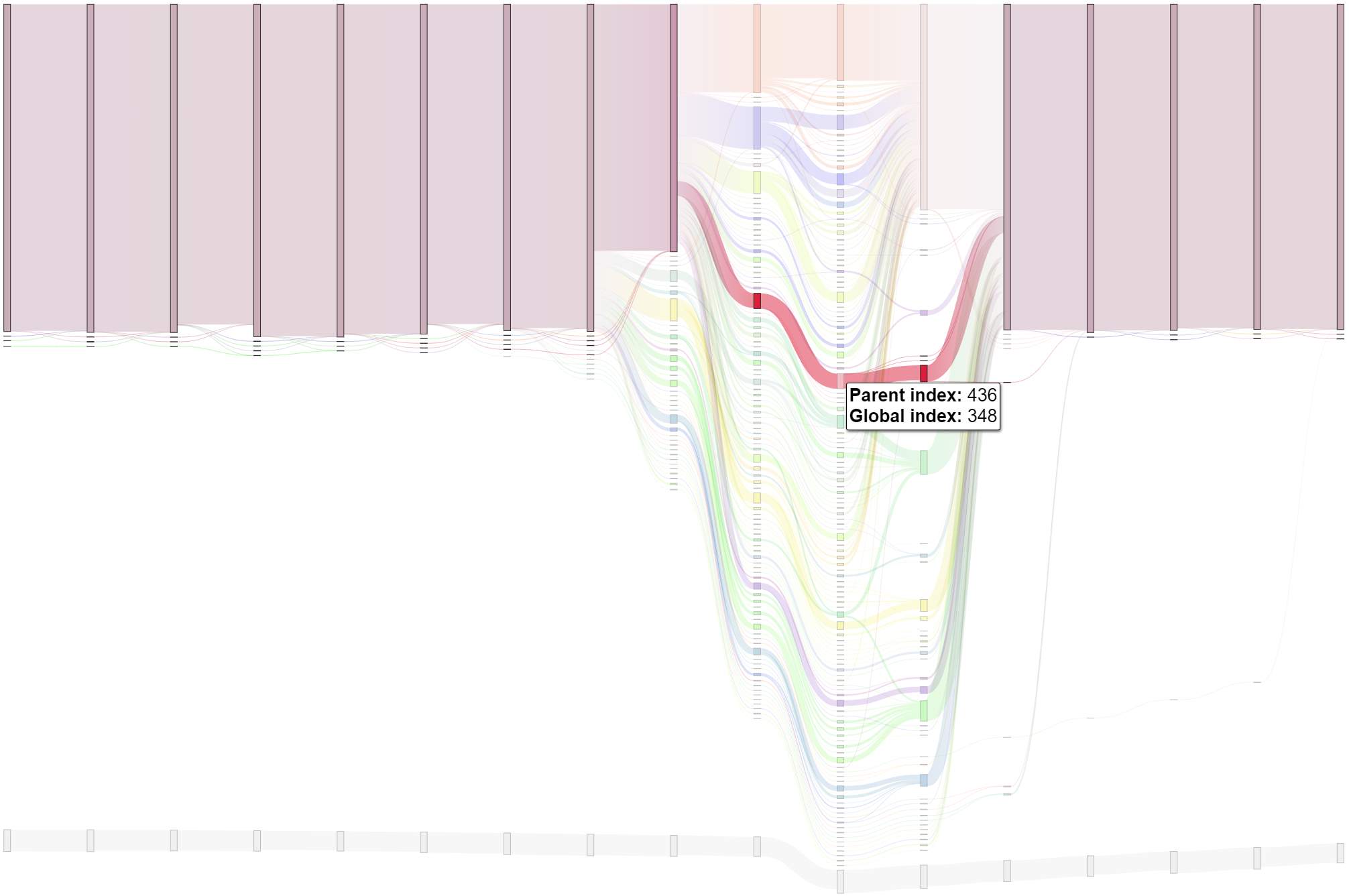}} 

    \caption{Experiment A: Hovering over segments' respective links in the tracking graph highlights the development of the selected (here red) region during the jamming phase, providing the history and future of the segment.}
    \label{fig:colormap_highlight}
\end{figure}

An assumption from the domain scientists is that the disks in the systems tend to preserve the history of the contacts. A consequence of this assumption would be that one can observe a high symmetry in the tracking graph during the loading and unloading phase, meaning that the splitting of the segments during the transition phase to jamming should be of the same nature as the merging of the segments during the unjamming phase. 
The capability to interactivity highlights the history and future of the node and the effectiveness of spatial colormap supports the comparison of patterns within the same subdomain~\cref{fig:colormap_highlight}.
Partially we can see such symmetries in the data, however not to the expected extent. In~\cref{fig:s3} some symmetries can be observed. E.g. the blue region is the last region that undergoes jamming and is also the first one undergoing unjamming. Also, the spatial representations of the regions seen at the bottom of the figure show some symmetries.
Also in~\cref{fig:teaser}, the green segments split earlier and merge later compared to the segments colored by blue and yellow that split later and merge earlier. However, overall the system during loading and unloading seems to be less symmetric in its details.


\begin{figure*}
    \centering
    {\includegraphics[width=\textwidth]{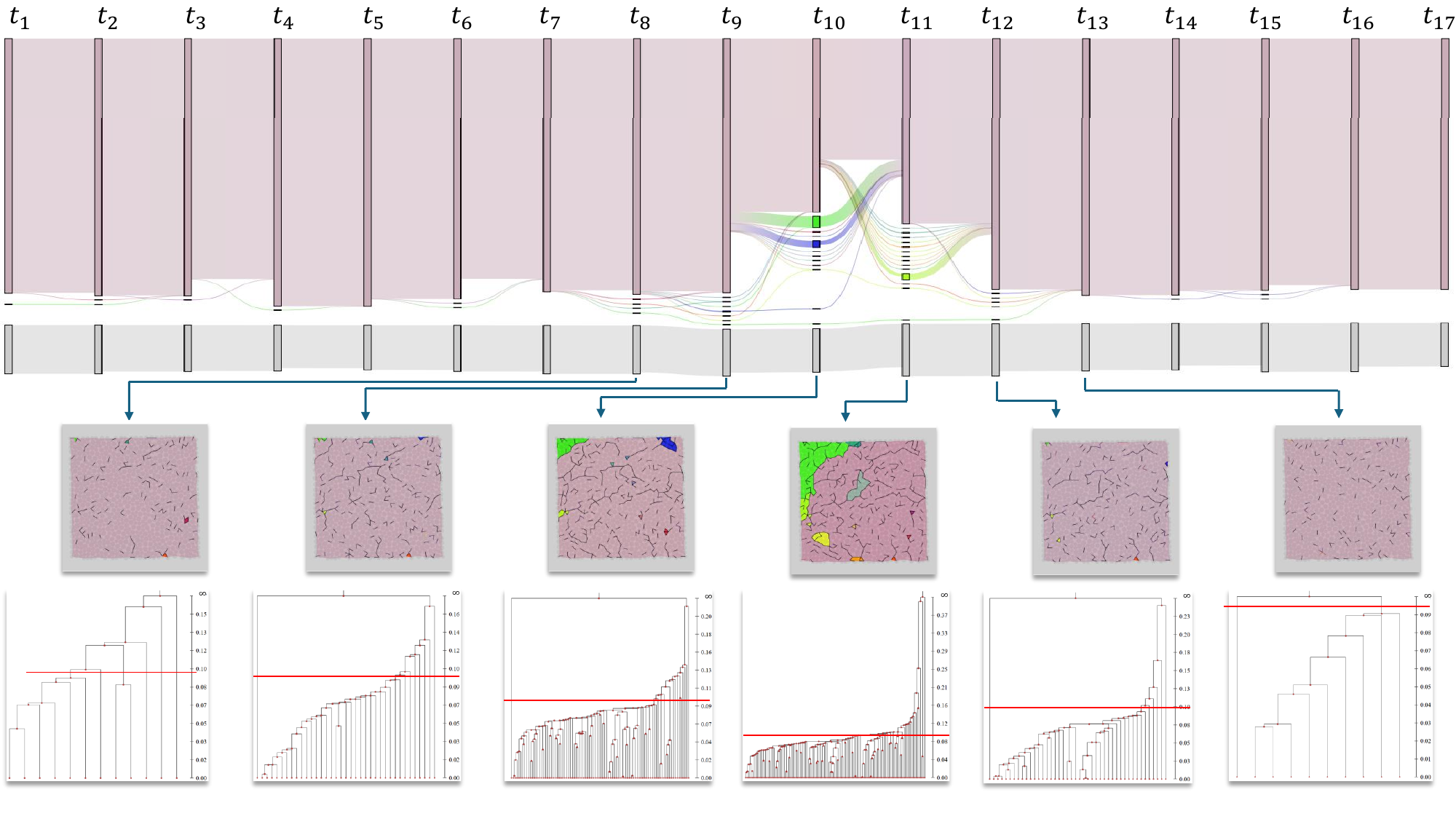}} 
    \caption{Experiment A: This image shows the tracking graph from~\cref{fig:s3} at a different hierarchy level. Here, the median of the force values at each time step is chosen as the filtration level.
    On the bottom, the cycle trees for the time steps during jamming are displayed. The red bars represent the filtration level. Please note that the trees have been cropped to only show the part of the tree with at least two cycles.}
    \label{fig:s7}
\end{figure*}

\section{Discussion and conclusion}

In this paper, we have demonstrated how cycle-tracking graphs can support the analysis of experiments involving two-dimensional granular materials. Representing an entire experiment as tracking graphs provides a compact visualization of the complex time series of force networks. By projecting individual time steps onto a one-dimensional axis, our method facilitates local analysis, while the connections between time steps offer a temporal perspective. This approach allows for observing the evolution of local patterns over time, inspecting symmetries in patterns, and providing an overview of the entire system, which can also be used to compare multiple experiments. Using node size and color in the tracking graph, based on the size and location of segments and hierarchy level, has proven very useful in our analysis. In future work, we plan to extend the use of these parameters to represent other derived attributes of the system.
We have proposed two different layouts for the tracking graph, each with its own advantages. The nested tracking graph provides an overview of the data at different levels simultaneously but can quickly become cluttered, making it difficult to follow individual segments. This representation does not support the analysis of a single time step. In contrast, the Sankey diagram scales well with the number of nodes and the complexity of the tracking graph. It supports both the analysis of a single time step and the overall evolution of the system. However, it currently does not support the direct display of the nesting hierarchy, which is a task we plan to address in future work.
\new{The overall tracking graph representation provides a good summary of the entire 2D photoelastic experiment for relatively small-scale experiments, such as those we are dealing with. However, this method does not scale well for experiments involving a large number of disks in terms of layout space. Another limitation of the presented method is that the concept of cycles and their tracking cannot directly translate to 3D granular materials.}

From a methodological perspective, we also foresee other application areas where tracking spatial partitioning or cycles could be beneficial, such as identifying non-coverage regions in dynamic sensor planar networks~\cite{dlotko2012distributed}.

\acknowledgments{
\new{This work was supported by the Wallenberg AI, Autonomous Systems and Software Program (WASP) funded by the Knut and Alice Wallenberg Foundation, the SeRC (Swedish e-Science Research Center) and the ELLIIT environment for strategic research in Sweden, the Swedish Research Council~(VR) grants 2019-05487 and 2023-04806, and an Indo-Swedish joint network project: DST/INT/SWD/VR/P-02/2019 VR grant 2018-07085.}}

\clearpage
\newpage
\bibliographystyle{abbrv-doi}

\bibliography{main}
\end{document}